\documentclass[12pt,a4paper]{article}
\usepackage{amsmath,amsfonts,amsthm,graphicx,lscape}
\usepackage[dvips]{psfrag}
\usepackage{cite}
\usepackage[normalem]{ulem}

\usepackage{textcomp}
\usepackage{amsmath,amsthm,amssymb,mathrsfs,cases}
\usepackage{amsfonts,graphicx,lscape}

\usepackage{accents}
\makeatletter
\def\widebar{\accentset{{\cc@style\underline{\mskip10mu}}}}
\makeatother

\theoremstyle{definition}

\begin{document}
\title{
Comment on 
``Discretisations of constrained KP hierarchies"
}
\author{Takayuki \textsc{Tsuchida}
}
\maketitle
\begin{abstract} 
In the recent paper (R.\ 
Willox 
and 
M.\ 
Hattori, arXiv:1406.5828), 
an integrable discretization of the 
nonlinear Schr\"odinger 
(NLS) 
equation 
is studied, 
which, they think, 
was 
discovered by
Date, Jimbo and Miwa in 1983 
and has been completely forgotten over the years. 
In fact, 
this 
discrete NLS 
hierarchy 
can be 
directly 
obtained from an elementary 
auto-B\"acklund transformation 
for 
the continuous NLS hierarchy 
and 
has been 
known since 1982. 
Nevertheless, 
it 
has been rediscovered 
again and again 
in the literature without attribution, 
so 
we 
consider 
it 
meaningful 
to 
mention 
overlooked 
original references  
on this 
discrete NLS hierarchy. 
%

\end{abstract}
%

%
%
\newpage
\noindent

The pioneering 
works of 
Calogero and Degasperis~\cite{Calo75,Calo76}, 
Chiu and Ladik~\cite{Chiu77},  
Hirota~\cite{Hiro77}
and 
Orfanidis~\cite{Orfa1,Orfa2}
in the late 1970s 
revealed  
that 
a certain class of auto-B\"acklund transformations 
and the associated nonlinear superposition principle 
(Bianchi's permutability theorem) 
can directly provide integrable discretizations
of the original continuous equations.  
Some relevant results 
can be found
in \cite{QNCL,NCWQ,NQC83}.

In 1982, 
Konopelchenko~\cite{Kono82} 
(also see 
Chudnovsky--Chudnovsky~\cite{Chud1,Chud2}) 
presented 
an elementary auto-B\"acklund transformation 
and the associated 
nonlinear superposition principle for the 
nonlinear Schr\"odinger 
(NLS) hierarchy; 
the latter reads 
\begin{equation}
\left\{
\begin{array}{l}
\displaystyle 
q_{m+1,n+1}  = q_{m,n} - 
 \frac{(\mu_n - \nu_m)q_{m+1,n}}{1 + q_{m+1,n} r_{m,n+1} }, 
\\[5mm]
\displaystyle 
r_{m+1,n+1} = r_{m,n} + 
 \frac{(\mu_n - \nu_m)
 r_{m,n+1}}{1 + q_{m+1,n} r_{m,n+1} }, 
\end{array}
\right.
\label{superpo}
\end{equation}
where $\mu_n$ and $\nu_m$ are arbitrary 
B\"acklund parameters that can depend on one of the two discrete 
independent variables $m$ and $n$. 
It was rediscovered by Date, Jimbo and Miwa~\cite{DJM83} in 1983
as an integrable discrete NLS system. 
Note
that unlike the 
Ablowitz--Ladik
discretizations~\cite{Chiu77,AL1,AL77}, 
this 
discretization 
does not admit 
the complex conjugation reduction 
between 
$q_{m,n}$ and $r_{m,n}$, 
so 
it 
is not 
a proper discretization of the NLS equation. 

By construction, 
the fully discrete NLS system (\ref{superpo}) 
possesses an infinite set 
of 
higher 
symmetries 
in each lattice direction. 
For example, 
the 
first 
nontrivial 
symmetry 
in the $n$-direction 
is given by 
the elementary auto-B\"acklund transformation~\cite{Kono82,Chud1}: 
\begin{equation}
\left\{
\begin{array}{l}
\displaystyle
 \frac{\partial q_{n}}{\partial x} 
	= -q_{n+1}  - \mu_n q_n + q_n^2 r_{n+1}, 
\\[3mm]
\displaystyle
 \frac{\partial r_{n+1}}{\partial x} = r_n + \mu_n r_{n+1}
 - q_n r_{n+1}^2 , 
\end{array}
\right.
\label{x-deri}
\end{equation}
for any fixed value of $m$. 
Other polynomial 
symmetries in the $n$-direction~\cite{Chud1}
are obtained from 
the continuous NLS flows 
by 
eliminating the $x$-derivatives using (\ref{x-deri}). 
Note that, 
with a minor reformulation, 
(\ref{x-deri}) can generate an infinite set 
of continuous NLS flows~\cite{Vek02,DM06,DM-SIGMA}, 
so 
all the information on 
an infinite number of polynomial higher 
symmetries in the $n$-direction 
is encoded in 
one 
symmetry (\ref{x-deri}). 
By changing the notation as 
\mbox{$r_{n+1} \to p_n$}, \mbox{$x \to - t_1$}, 
we obtain 
a more familiar form of the 
differential-difference system~\cite{Chud1,Chud2}: 
\begin{equation}
\left\{
\begin{array}{l}
\displaystyle
 \frac{\partial q_{n}}{\partial t_1} 
	= q_{n+1} + \mu_n q_n - q_n^2 p_n, 
\\[3mm]
\displaystyle
 \frac{\partial p_{n}}{\partial t_1} = -p_{n-1} - \mu_n p_{n}
 + q_n p_{n}^2 . 
\end{array}
\right.
\label{x-deri2}
\end{equation}
This lattice system 
(mostly in the simple 
case of \mbox{$\mu_n = 
0$ or $-1$}) has been 
rediscovered repeatedly 
in the literature; 
some earlier references are \cite{SY91,SviYami91,Zhang91}
%
%
%
and 
the inverse scattering method was developed 
in~\cite{Bhate}. 

Even without such preknowledge about its 
origin,  
the 
fully 
discrete 
system (\ref{superpo}) 
allows us to generate 
an infinite set of higher symmetries 
using the notion of Miwa shifts. 
To this aim, 
we 
consider a rescaling 
\mbox{$q_{m+1,n} \to - \frac{1}{\nu_m} q_{m+1,n}$},  
\mbox{$r_{m+1,n} \to - \nu_m r_{m+1,n}$} 
for all $n$ 
and rewrite (\ref{superpo}) as 
\begin{equation}
\left\{
\begin{array}{l}
\displaystyle 
q_{m+1,n+1}  = -\nu_m q_{m,n} - 
 \frac{(\mu_n - \nu_m)q_{m+1,n}}{1 - \frac{1}{\nu_m} q_{m+1,n} r_{m,n+1} }, 
\\[5mm]
\displaystyle 
- \nu_m r_{m+1,n+1} = r_{m,n} + 
 \frac{(\mu_n - \nu_m)
 r_{m,n+1}}{1 - \frac{1}{\nu_m} q_{m+1,n} r_{m,n+1} }. 
\end{array}
\right.
\label{superpo2}
\end{equation}
Then, by setting \mbox{$\nu_m = -\frac{1}{h_m}$} 
where $h_m$ is a step-size 
parameter 
in the $m$-direction and 
taking the limit \mbox{$h_m \to 0$}, we 
obtain higher symmetries starting from 
(\ref{x-deri}). 

Besides such polynomial higher symmetries, 
the fully discrete NLS system (\ref{superpo}) also 
possesses rational 
higher 
symmetries. 
By setting \mbox{$\mu_n=\mu $}, 
\mbox{$\nu_m = \mu + \delta_m $} 
where $\delta_m$ is a step-size 
parameter 
in the \mbox{$(m+n)$}-direction and 
taking the limit 
\mbox{$\delta_m \to 0$} 
in such a way that \mbox{$q_{m+1,n+1}  - q_{m,n} \to 0$}, we 
obtain higher 
symmetries in the $n$-direction 
starting from 
\begin{equation}
\left\{
\begin{array}{l}
\displaystyle 
\frac{\partial q_n}{\partial t_{-1}} =  
 \frac{q_{n-1}}{1 + q_{n-1} r_{n+1} }, 
\\[5mm]
\displaystyle 
\frac{\partial r_n}{\partial t_{-1}} = - 
 \frac{
 r_{n+1}}{1 + q_{n-1} r_{n+1} }.
\end{array}
\right.
\label{superpo3}
\end{equation}
By changing the notation as 
\mbox{$r_{n+1} \to p_n$}
and superimposing  (\ref{x-deri2}) 
in the case of a constant \mbox{$\mu_n 
$}, 
we obtain 
a (not proper) integrable 
space discretization of the NLS system: 
\begin{equation}
\left\{
\begin{array}{l}
\displaystyle
 \frac{\partial q_{n}}{\partial t} 
	= a ( q_{n+1}  - q_n^2 p_n ) 
	+ b q_n + c \frac{q_{n-1}}{1 + q_{n-1} p_{n} }, 
\\[3mm]
\displaystyle
 \frac{\partial p_{n}}{\partial t} = -a ( p_{n-1} 
 - q_n p_{n}^2) - b p_{n} - c
 \frac{
 p_{n+1}}{1 + q_{n} p_{n+1} }. 
\end{array}
\right.
\label{x-deri3}
\end{equation}
This lattice system was 
introduced 
by Gerdjikov and Ivanov~\cite{GI2} in 1982 
and rediscovered by Merola, Ragnisco and Tu~\cite{Tu}; 
it 
is related to the Ablowitz--Ladik lattice~\cite{AL1} 
through a nonlocal transformation of dependent variables 
involving an infinite product~\cite{GI2}. This transformation 
can be rewritten in a local form if one uses 
a $\tau$-function formalism 
(cf.~\cite{TD11}).
Note incidentally that 
in arXiv:1204.2928,  
Gerdjikov proposed 
a new type of integrable three-wave 
equations with a Lax pair, 
but 
these 
results 
were 
previously 
reported 
in 
\S3.3 of 
arXiv:1012.2458. 

All the above discussion (except 
the $\tau$-function formalism) can be generalized 
straightforwardly 
to the case of 
matrix-valued dependent variables, but we focused 
on 
the scalar 
case 
to enhance readability.

\end{document}